\documentstyle[aps,prl,multicol,floats,epsf]{revtex}

\begin{document}
\draft
\wideabs{

\title{Carrier concentrations in Bi$_{2}$Sr$_{2-z}$La$_z$CuO$_{6+\delta}$ 
single crystals and \\
their relation to Hall coefficient and thermopower}

\author{Yoichi Ando,$^{1,2}$ Y. Hanaki,$^{1,2}$ S. Ono,$^{1}$ 
T. Murayama,$^{1,2,}$\cite{Mura} Kouji Segawa,$^{1}$ 
N. Miyamoto,$^{1,2}$ and Seiki Komiya$^{1}$}
\address{$^1$Central Research Institute of Electric Power
Industry, Komae, Tokyo 201-8511, Japan}
\address{$^{\rm 2}$ Department of Physics, Science University of Tokyo, 
Shinjuku-ku, Tokyo 162-8601, Japan}

\date{2201conc-3.tex}
\maketitle

\begin{abstract}
We measured the thermopower $S$ and the Hall coefficients $R_H$ of 
Bi$_{2}$Sr$_{2-z}$La$_z$CuO$_{6+\delta}$ (BSLCO) 
single crystals in a wide doping range, in an effort to 
identify the actual hole concentrations per Cu, $p$, 
in this system. 
It is found that the ``universal" relation between the 
room-temperature thermopower and $T_c$ does not hold in the BSLCO system.
Instead, comparison of the temperature-dependent $R_H$ data with 
other cuprate systems is used as a tool to identify the 
actual $p$ value.  To justify this approach, we compare normalized 
$R_H(T)$ data of BSLCO, La$_{2-x}$Sr$_{x}$CuO$_{4}$ (LSCO), 
YBa$_2$Cu$_3$O$_{y}$, and Tl$_{2}$Ba$_{2}$CuO$_{6+\delta}$,
and demonstrate that the $R_H(T)$ data of the LSCO system can be
used as a template for the estimation of $p$.
The resulting phase diagram of $p$ vs $T_c$ suggests that $T_c$ is 
anomalously suppressed in the underdoped samples, becoming zero at 
around $p \simeq$ 0.10, while the optimum $T_c$ is achieved at 
$p \simeq$ 0.16 as expected.
\end{abstract}


}
\narrowtext

Determination of the actual carrier concentration in the 
high-$T_c$ cuprates is in general a difficult task.  
The La-214 system [La$_{2-x}$Sr$_{x}$CuO$_{4}$ (LSCO) or
La$_{2-x}$Ba$_{x}$CuO$_{4}$ (LBCO)] is almost the only system 
where the carrier concentration is nearly unambiguously known; 
in this system, the hole concentrations per Cu, $p$, is identical 
to the $x$ value, as long as the oxygen is stoichiometric.
In the YBa$_2$Cu$_3$O$_{y}$ (YBCO) system, the hole concentration 
in the CuO$_2$ planes is ambiguous because of the existence of the 
imperfect CuO chains which absorb part of the doped holes.
In other systems like Bi$_{2}$Sr$_{2}$CaCu$_{2}$O$_{8+\delta}$ 
(Bi-2212), Bi$_{2}$Sr$_{2}$CuO$_{6+\delta}$ (Bi-2201), or 
Tl$_{2}$Ba$_{2}$CuO$_{6+\delta}$ (Tl-2201), 
the determination of the hole concentration is also ambiguous 
because Bi and Tl ions have mixed valencies \cite{Idemoto}.

In ordinary metals or semiconductors, the Hall coefficient $R_H$ is 
often used for the determination of the carrier concentration.
However, $R_H$ of the cuprates has not been considered
to be a useful tool to determine $p$, 
because $R_H$ shows a rather strong temperature dependence.
Moreover, it has been reported for LSCO that the 
magnitude of $R_H$ is several times smaller than that expected from 
the chemically-determined carrier concentration \cite{Takagi}.
On the other hand, it has been proposed \cite{Tallon} 
that the magnitude of the thermopower at room temperature (290 K), 
denoted as $S(290{\rm K})$, can be used for the determination of $p$,
based on the observation that the plot of $T_c/T_c^{max}$ vs $S(290{\rm K})$ 
is almost ``universal" among many cuprates ($T_c^{max}$ 
is the optimum $T_c$ of each system), although the data of the 
LSCO system do not follow the ``universal" relation \cite{Cooper}.
If one assumes another ``universal" relation \cite{Presland} 
between $p$ and $T_c/T_c^{max}$, so-called the ``bell shape", 
the measurement of $S(290{\rm K})$ yields an 
estimation of $p$ as long as the two ``universal" relations hold.

The above mentioned relation between $p$ and $S(290{\rm K})$ has not been 
tested in Bi$_{2}$Sr$_{2-z}$La$_z$CuO$_{6+\delta}$ 
(BSLCO, or La-doped Bi2201), in which the carrier concentration can be 
changed over a wide range \cite{Murayama}. 
An increase in the La concentration $z$ in this system leads to a 
smaller density of holes in the CuO$_2$ planes, and the optimum $T_c$ 
is achieved with $z \simeq$ 0.4.
In this paper, we report our systematic measurements of $S(290{\rm K})$ 
and $R_H(T)$ for a series of BSLCO single crystals, 
for which $z$ is varied from 0.2 to 1.0.  
It is found that the optimally-doped BSLCO crystal
show notably smaller $S(290{\rm K})$ value than that expected 
from the ``universal" relation.  This is a strong indication 
that the ``universal" relation does not hold in the BSLCO system
and thus one should not use this relation to determine $p$.
We therefore tried to use the temperature-dependent $R_H$ for the 
determination of $p$.  
We show that, using $R_H(T)$ of the LSCO system as a template, 
the $R_H(T)$ data can give reasonable estimate of $p$. 
With the $p$ values thus obtained, we construct the relation between 
$p$ and $S(290{\rm K})$, which is actually quite different from the 
``universal" relation.
The final phase diagram of $T_c$ vs $p$ for the BSLCO system suggests that 
$T_c$ is anomalously suppressed in the underdoped samples, becoming zero 
at around $p \simeq$ 0.10, while the optimum $T_c$ is achieved at 
$p \simeq$ 0.16.

The single crystals of Bi$_{2}$Sr$_{2-z}$La$_z$CuO$_{6+\delta}$ 
are grown using a floating-zone (FZ) technique \cite{Murayama}.  
Here we report crystals with $z$ from 0.23 to 1.02.
The crystal with $z$=0.39 is optimally-doped;
the optimum $T_c$ is 36 K, which is very high 
for BSLCO \cite{Murayama} and indicates that the crystals reported here 
are among the best crystals available to date.
The actual La concentration in the crystals are 
determined by employing both the inductively-coupled plasma 
analysis and the electron-probe microanalysis.
For the transport measurements, the crystals are cut into dimensions 
of typically 2 $\times$ 1 $\times$ 0.05 mm$^{3}$. 
Since the absolute magnitude of the Hall coefficient is important for 
this work, the thickness of the samples is accurately determined by 
measuring their weight with 0.1-$\mu$g resolution, and 
the uncertainty in the absolute magnitude of $R_H$ is estimated to be 
less than $\pm$8\%.
All the crystals are annealed in flowing oxygen at 
650$^\circ$C for 48 hours to guarantee uniform oxygen distribution 
in the samples.

We define $T_c$ in this paper by the onset of the Meissner effect 
measured by dc magnetic susceptibility.  We have confirmed that 
the Meissner-onset $T_c$ agrees very well with the zero-resistance $T_c$ 
for our crystals \cite{Murayama}.
The Hall coefficient $R_H$ is measured together with the in-plane resistivity 
$\rho_{ab}$ by using a standard six probe technique.  
The current contacts are carefully painted to cover two opposing 
side faces of the platelet-shaped crystals to ensure a uniform current flow.  
The voltage contacts are painted on the two remaining side faces of the 
crystals (not on the top or bottom faces), 
which is important for the accurate Hall effect measurement.
The Hall coefficients are measured by sweeping the magnetic field to 
both plus and minus polarities, while the temperature is stabilized by 
a combination of the calibrated Cernox sensor and a capacitance sensor 
\cite{Murayama}.
The thermopower is measured with a standard steady-state technique 
with a reversible temperature gradient of $\sim$1 K, and the thermopower 
of the gold wires are corrected for.

\begin{figure}[t!]
\epsfxsize=0.75\columnwidth
\vspace{0.5cm}
\centerline{\epsffile{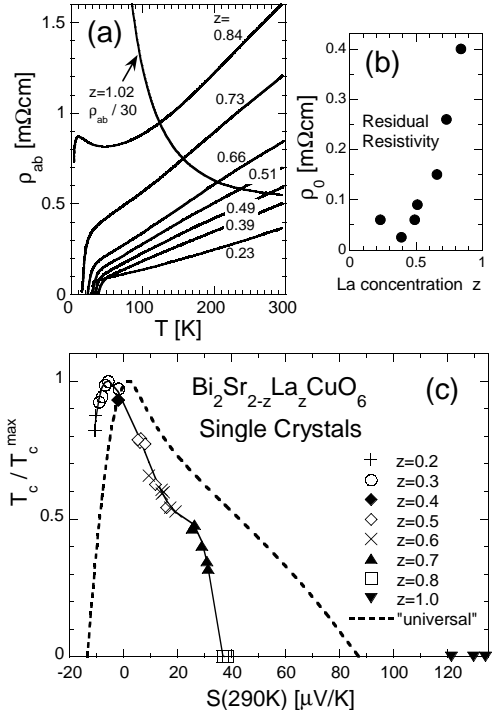}}
\vspace{0.5cm}
\caption{
(a) $\rho_{ab}(T)$ data of the BSLCO crystals for selected $z$ values.
(b) $z$ dependence of $\rho_0$ obtained from Fig. 1(a).
(c) $T_c/T_c^{max}$ vs $S(290{\rm K})$ of the BSLCO crystals, 
plotted together with the ``universal" 
relation \protect\cite{Tallon} reproduced as a dashed line.
The $z$ values shown in this panel are nominal values.
The solid line is a guide to the eyes.}
\label{fig1}
\end{figure}

Figure 1(a) shows the $\rho_{ab}(T)$ data of the BSLCO crystals 
for selected $z$ values from 0.23 - 1.02, which show systematic
evolution with changing carrier concentration.
One may notice that the residual resistivity $\rho_0$ of these samples
\cite{rho0} becomes systematically larger with increasing La doping, 
although $\rho_0$ is very small ($\sim$20 $\mu\Omega$cm) at 
optimum doping; the $z$ dependence of $\rho_0$ is plotted in Fig. 1(b). 
Figure 1(c) shows the plot of $T_c/T_c^{max}$ vs $S(290{\rm K})$ of our 
BSLCO crystals ($T_c^{max}$ = 36 K), together with the ``universal" 
relation \cite{Tallon} reproduced as a dashed line. 
It is clear that the two curves do not agree at all. 
This is either because the universal relation between $p$ and 
$S(290{\rm K})$ does not hold in the BSLCO system or 
because the $T_c$ values are somehow reduced from the ideal value 
(and thus does not correctly reflect the hole concentrations).
To determine which is actually the case, we pay attention to 
the peaks in Fig. 1(c), which corresponds to the optimum doping;
if we can find an evidence from another experiment that the 
optimum $T_c$ of BSLCO is indeed realized at $p \simeq$ 0.16, 
it is a clear indication that the universal relation of $T_c/T_c^{max}$ 
vs $S(290{\rm K})$ is disobeyed in BSLCO.
Note that the $S(290{\rm K})$ value of the optimally-doped BSLCO sample 
($\sim -6$ $\mu$V/K) corresponds to $p \simeq$ 0.22 in the 
``universal" relation \cite{Tallon}.

\begin{figure}[t!]
\epsfxsize=0.8\columnwidth
\centerline{\epsffile{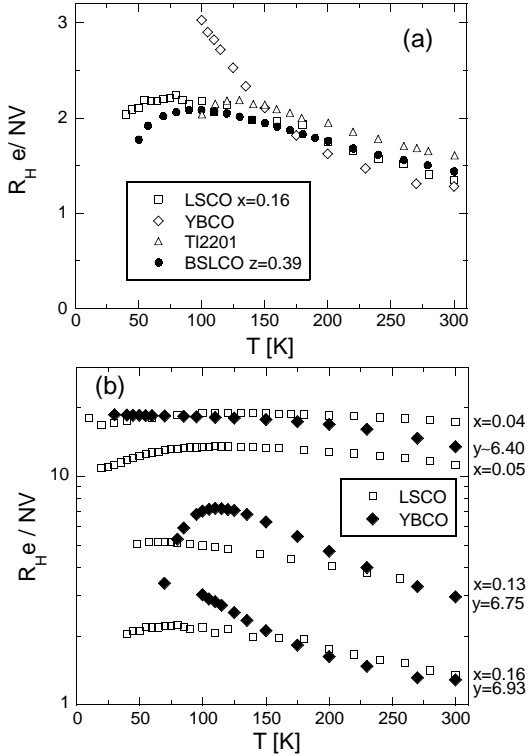}}
\vspace{0.5cm}
\caption{
(a) Plots of $R_He/NV$ vs $T$ for various optimally-doped cuprates, 
LSCO ($x$=0.16), YBCO ($y$=6.93), Tl-2201, and BSLCO ($z$=0.39). 
The data for Tl-2201 is taken from Ref. \protect\cite{Kubo}.
Note that all the data lie in a rather narrow band of $\sim$20\% in the 
temperature range 150 - 300 K, implying that $R_He/NV$ can be a 
good measure of the hole concentration.
(b) Comparison of $R_He/NV$ of LSCO and YBCO
at three representative dopings: optimum doping, 1/8 doping, and the 
superconductor-insulator boundary ($p \simeq$ 0.05).}
\label{fig2}
\end{figure}

We now demonstrate that the magnitude of the Hall coefficient 
can be used as a guide to estimate $p$.  
It has been pointed out \cite{Cooper} that the fictitious carrier density 
calculated from the room-temperature value of $R_H$ does not vary much 
among different cuprates at the same hole concentration.  
Therefore, if $R_H$ is normalized by the unit volume $V$ and the number of 
Cu atoms in the unit, $N$, the resulting $R_H/NV$ is expected to show 
similar values for different cuprates at optimum doping ($p \simeq$ 0.16).
Note that $R_He/NV$ ($e$ is the electronic charge) gives the inverse of 
the fictitious hole density per Cu.  
Figure 2(a) shows the plot of $R_He/NV$ vs $T$ for various cuprates, 
LSCO, YBCO, Tl-2201, and BSLCO, at optimum doping. 
The data for Tl-2201 is taken from Ref. \cite{Kubo}, and all other data 
are measured by ourselves paying particular attention to the accuracy in 
the absolute magnitude.
The LSCO sample is a high-purity polycrystal \cite{Miyamoto}, and 
the YBCO sample is a high-quality single crystal grown by a flux 
method using Y$_2$O$_3$ crucible \cite{Segawa}. 
In the $R_H$ data of LSCO and YBCO, the uncertainty in the 
absolute magnitude is less than $\pm$8\%. 
It is clear in Fig. 2(a) that the normalized $R_H$ of all the optimally-doped 
cuprates shown here agree reasonably well in the temperature range 150 - 300 K, 
where the data fall in a rather narrow band of $\sim$20\%. 
This observation is remarkable and has not been emphasized before in 
the literature.
For the purpose of this paper, Fig. 2(a) strongly suggests 
that it is reasonable to assert $p$ to be actually around 0.16 
in our optimally-doped BSLCO crystal. 

Figure 2(b) shows the comparison of $R_He/NV$ vs $T$ for LSCO and YBCO
at three representative dopings: optimum doping, 1/8 doping, and the 
superconductor-insulator (S-I) boundary ($p \simeq$ 0.05).  
For YBCO, it has been reported \cite{Tallon2} that 
the 60-K phase ($y \simeq$ 6.7) corresponds roughly to $p \simeq$ 1/8 
and the S-I boundary ($y \simeq$ 6.4) lies at 
$p \simeq$ 0.05.  Therefore, because of the special physical meanings 
attached to these hole concentrations, the three dopings shown in Fig. 2(b) 
give good reference points where the $p$ values are well-defined.
One can see in Fig. 2(b) that the $R_He/NV$ data of the two systems agree 
reasonably well at the three dopings at temperatures above $\sim$200 K.  
This observation gives 
further support to the idea that $R_He/NV$ can be used as a guide to 
estimate $p$; the comparison shown in Fig. 2(b) suggests that 
the data of LSCO system, for which $p$ is nearly unambiguous, can be used 
as a template to compare the data of other systems.

\begin{figure}[t!]
\epsfxsize=0.8\columnwidth
\centerline{\epsffile{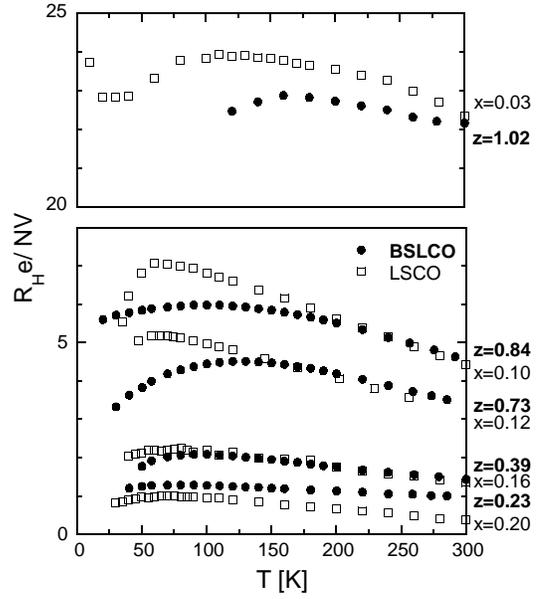}}
\vspace{0.5cm}
\caption{
Comparison of $R_H(T)e/NV$ of LSCO and BSLCO
at various hole concentrations.}
\label{fig3}
\end{figure}

Based on the above observation, we try to estimate $p$ of our BSLCO
crystals by comparing the $R_He/NV$ data with that of LSCO.
Figure 3 shows the data of the two systems for selected concentrations.
From Fig. 3, one can infer that the La concentration, $z$, of 0.73 corresponds 
to $p \simeq$ 0.12, $z$=0.84 to $p \simeq$ 0.10, and $z$=1.0 to 
$p \simeq$ 0.03.
Based on this observation, we determine the $p$ values for various $z$ 
as follows:
$p$ = 0.18, 0.16, 0.15, 0.14, 0.13, 0.12, 0.10, and 0.03, for 
$z$ = 0.23, 0.39, 0.49, 0.51, 0.66, 0.73, 0.84, and 1.02, respectively
\cite{note}.
Note that these determination are not precise and the expected accuracy
would be $\pm$10\% at best \cite{note2}.
This result implies that the superconductivity of our BSLCO is about to 
disappear at $z$=0.84 {\it not} because the hole concentration is reduced 
to $p \sim$ 0.05.
Remember that there is a tendency that $\rho_0$ becomes 
systematically larger with increasing La doping, as shown in Fig. 1(b).
This suggests that one possible reason for the disappearance of 
superconductivity at $z \simeq$ 0.9 (which corresponds to 
$p$ = 0.09 - 0.10) is the increased disorder in the heavily-La-doped crystals.
In other words, in our BSLCO samples the superconductivity is 
more significantly suppressed in more underdoped samples because of 
the larger amount of disorder probed by $\rho_0$.
The microscopic origin of this disorder triggered by heavy La doping 
is yet to be elucidated; it is difficult to imagine that the random
potential produced by the replacement of Sr with La itself is 
responsible to produce $\rho_0$ of as large as 400 $\mu\Omega$cm.

\begin{figure}[t!]
\epsfxsize=0.8\columnwidth
\centerline{\epsffile{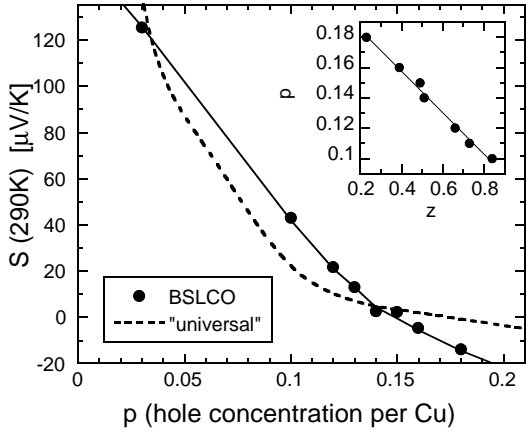}}
\vspace{0.5cm}
\caption{
$S(290{\rm K})$ vs $p$ plot for the BSLCO crystals
together with the ``universal" relation \protect\cite{Tallon}.
The solid line is a guide to the eyes.
Inset shows a plot of estimated $p$ vs $z$, together with a 
straight line fit.}
\label{fig4}
\end{figure}

Figure 4 shows the plot of $S(290{\rm K})$ vs $p$ for our BSLCO crystals
(where $p$ is determined as above), together with the 
``universal" relation that is reported to be valid for most of the 
cuprates except for LSCO \cite{Cooper}.  
Clearly, the BSLCO system does not follow the ``universal" trend.  
The reason for this deviation is not clear, but 
it is intriguing to note that both BSLCO and LSCO are peculiar systems 
in which the optimum $T_c$ is much lower than the ``intrinsic" $T_c$ 
(about 80 K) expected for single-layer cuprates.  
It has been discussed \cite{Baskaran} that these two systems 
might share the common trend to stabilize charged stripes, which is known 
to suppress superconductivity \cite{Tranquada}.

Using the $p$ values inferred for our BSLCO samples,
we can estimate the fictitious $T_c^0$ values for each La 
concentration by assuming the ordinary ``bell shape" \cite{Presland}
for $T_c^0/T_c^{max}$ vs $p$; this $T_c^0$ is the ideal value that would 
be expected for a given $p$ if there were no cause for the reduction of 
$T_c$.
Since $\rho_0$ of the $z$=0.39 sample 
is small [Fig. 1(b)], we assume that our optimum $T_c$ is not 
significantly affected by disorder and thus we take $T_c^{max}$ = 36 K
for the calculation of $T_c^0$.
Figure 5 shows the plot of actual $T_c$ vs $p$ of the BSLCO system,
together with the fictitious $T_c^0$ vs $p$.
It is clear in Fig. 5 that $T_c(p)$ of the BSLCO system shows a faster 
drop in both the underdoped and overdoped sides of the phase diagram, 
which is probably caused by disorder as inferred from the increase in 
$\rho_0$ when $z$ moves away from the optimum doping [Fig. 1(b)].

In summary, we estimate the hole concentration per Cu, $p$, of a 
series of BSLCO crystals by using the normalized Hall coefficient 
$R_He/NV$ and its comparison with other cuprates. 
It is demonstrated that at optimum doping $R_H(T)e/NV$ of various 
cuprates (LSCO, YBCO, Tl-2201, and BSLCO) 
agree reasonably well in the temperature range 150 - 300 K, 
where the data fall in a rather narrow band of $\sim$20\%.
This implies that $R_He/NV$ can be used as a guide to estimate $p$.
It is found that in the BSLCO system the room-temperature 
thermopower $S(290{\rm K})$ as a function of $p$ does not follow the 
``universal" trend.
The phase diagram of $T_c$ vs $p$ for our BSLCO crystals suggests that 
$T_c$ is anomalously suppressed in the underdoped samples, 
where the residual resistivity is found to increase systematically
with increasing La doping.

\begin{figure}[t!]
\epsfxsize=0.8\columnwidth
\centerline{\epsffile{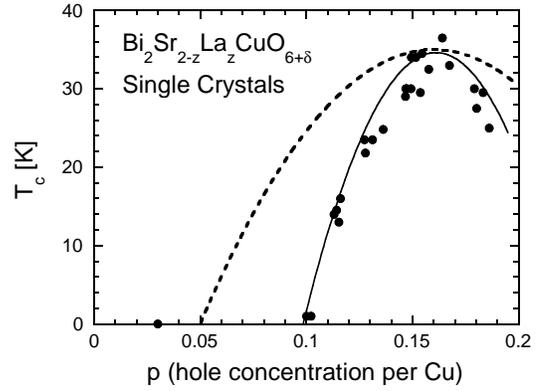}}
\vspace{0.5cm}
\caption{
Actual $T_c$ (solid circles) and the fictitious $T_c^0$ 
(dashed line) plotted vs $p$ for the BSLCO crystals. 
The $p$ value is calculated from $z$ by using $p = 0.21 - 0.13z$
\protect\cite{note2}.
The solid line is a guide to the eyes.}
\label{fig5}
\end{figure}

%
\medskip
\vfil


\begin{references}
\bibitem{rho0}
$\rho_0$ is obtained by extrapolating the $T$-linear part of 
$\rho_{ab}(T)$ to 0 K, except for the $z$=0.23 sample for which 
$\rho_{ab}(T)$ is well fitted by $\rho_0 + aT^{1.3}$.
\bibitem{note}
It is intriguing that $p$ changes drastically between $z$=0.84 and 1.02,
where the superconductivity is lost.  
It is possible that a strong localization of the carriers (which is known 
from the $\rho_{ab}(T)$ data for $z$=1.02) is responsible for the loss of 
mobile carriers.
\bibitem{note2}
For a limited range of $0.23 \le z \le 0.84$, 
the obtained relation between $p$ and $z$ 
can be approximated by a linear relation $p = 0.21 - 0.13z$,
see inset to Fig. 4.
\end{references}
\end{document}